\documentclass{iau} 
\usepackage{graphicx}
\usepackage[version=4]{mhchem}

\title[Molecules in PNe] 
{Molecular studies of Planetary Nebulae}

\author[Y. Zhang]
{Yong Zhang$^{1,2}$}

\affiliation{$^1$Department of Physics, The University of Hong Kong, Pokfulam Road, Hong Kong, China\\
$^2$  Laboratory for Space Research, Faculty of Science, The University of Hong Kong, Pokfulam Road, Hong Kong, China
\\email: {\tt zhangy96@hku.hk}}

\pubyear{2016}
\volume{323}  
\jname{Planetary Nebulae: Multi-wavelength Probes of Stellar and Galactic Evolution}
\editors{X.W. Liu, L. Stanghellini \& A. Karakas, eds.}
\begin{document}

\maketitle

\begin{abstract}
Circumstellar envelopes (CEs) around evolved stars are an active site for the production of molecules. After evolving through the Asymptotic Giant Branch (AGB), proto-planetary nebula (PPN), to planetary nebula (PN) phases, CEs ultimately merge with the interstellar medium (ISM). The study of molecules in PNe, therefore,
is essential to understanding the transition from stellar to interstellar 
materials. So far, over 20 molecular species have been discovered in PNe.
The molecular composition of PNe is rather different from those of AGB
and PPNe, suggesting that the molecules synthesized in  PN progenitors have
been heavily processed by strong ultraviolet radiation from the central star.
Intriguingly, fullerenes and complex organic compounds having
aromatic and aliphatic structures can be rapidly formed and largely survive
during the PPN/PN evolution. The similar molecular compositions in PNe
and diffuse clouds as well as the detection of C$_{60}^+$ in the ISM 
reinforce the view that the mass-loss from PNe can significantly enrich the 
ISM with molecular species, some of which may be responsible for the
diffuse interstellar bands.
In this contribution, I briefly summarize some recent observations of molecules in PNe, with 
emphasis on their implications on circumstellar chemistry.
\end{abstract}

\firstsection 
\section{Introduction}

The mass loss of evolved stars with intermediate-mass progenitors
 creates an expanding circumstellar envelope (CE), in which 
a variety of molecules are synthesized. During the CE evolution from
the Asymptotic Giant Branch (AGB), proto-planetary nebula (PPN), to 
planetary nebula (PN) phases, the physical conditions dramatically
alter, resulting in considerable changes in molecular composition.
The yielded molecules eventually disperse into the interstellar medium (ISM), 
and may act as the seed of complex organic molecules discovered in molecular clouds
(\cite[Ziurys 2008]{zi08}). Therefore, the investigation of molecules in PNe provides essential
information on the material cycle in galaxies. The very existence 
of molecular gas in PNe can be inferred from the large discrepancy between the
mass of the PN progenitor (1--8\,$M_\odot$) and the masses of PN central stars and 
ionized nebulae ($\sim0.9$\,$M_\odot$), known as the missing mass problem
(\cite[Kwok 1994]{kwok94}).

Up to now more than 80 different molecular species have been detected in C-rich AGB envelopes and PPNe.
Numerous models have been developed  to simulate the chemical processes in AGB envelopes
(\cite[see Millar 2016, for a recent review]{m16}). The standard picture is that
the parent molecules are formed through the equilibrium chemistry near the stellar photosphere,
and then accelerated by radiation pressure on dust into outer regions where more complex
molecules are synthesized by the chemistry induced by the external interstellar ultraviolet radiation 
field. The chemical evolution during the AGB-PPN-PN transitions
has been  investigated through  systematic molecular line surveys
(\cite[see Bujarrabal 2006, Cernicharo et al. 2011, for detailed reviews]{bu06,cer11}), which clearly
imply dramatic changes of molecular composition at the PPN phase 
(\cite[e.g. Pardo et al. 2007, Park et al. 2008, Zhang et al. 2013]{pardo07,park08,zhang13}).
The dynamical timescales of the three evolutionary stages can impose constraints on
the efficiencies of chemical reactions. Thus CEs of evolved stars can serve as a laboratory
for studying the chemical processes under extreme conditions.

\begin{table}
  \begin{center}
  \caption{The molecules detected in PNe.}
  \label{tab1}
 {\scriptsize
  \begin{tabular}{lllll}\hline
{\bf 2} & {\bf 3} & {\bf 4} & {\bf 5} & {\bf $>5$} \\
{\bf atoms} & {\bf atoms} & {\bf atoms} & {\bf atoms} & {\bf atoms} \\ \hline
H$_2$ & C$_2$H      & NH$_3$  & $c$-C$_3$H$_2$ & C$_{60}$  \\
CH    & HCN         & H$_2$CO &  HC$_3$N       & C$_{60}^+$\\
CH$^+$& HCO$^+$     &         &                & C$_{70}$  \\
CN    & H$_2$O      &         &                &           \\
CO    & HNC         &         &                &           \\
CO$^+$& N$_2$H$^+$  &         &                &           \\
OH    & SO$_2$      &         &                &           \\
OH$^+$& HCS$^+$?    &         &                &           \\
SiO   & SiC$_2$?    &         &                &           \\
CS    &             &         &                &           \\
SO    &             &         &                &           \\
NS    &             &         &                &           \\
\hline
  \end{tabular}
  }
 \end{center}
 \scriptsize{
 {\it Note:} Not include the molecular-rich pre-PN CRL\,618 and solid-phase species.
}
\end{table}

With the evolution to the PN phase, increasing central-star temperature 
 results in a photoionized region surrounded by a photon-dominated region (PDR), where 
 the chemistry is dominated by the photons of energy from 6 to 13.6\,eV, which 
contribute significantly to gas heating by ejecting electrons from dust grains and induce 
ion-neutral reactions. Shocks may play an important role in circumstellar chemistry
by inducing endothermic reactions and destructing dust into molecules, though strong shocks might destroy molecules. Although high photodissociation and dissociative recombination rates
render the environments of PNe  hostile for the survival of molecules, the ultraviolet 
radiation can be largely attenuated by clumps, density-enhanced tori, as well as self- and 
mutual-shielding of molecules. These optically thick regions and  the 
extensive gas envelopes presumably have different chemical evolution routes from one another.  The first chemical model of PNe
was developed by \cite[Black (1978)]{black78} who predicted the existence of simple molecules such as
H$_2$, H$_2^+$, HeH$^+$, OH, and CH$^+$ in PDRs. Although most of them were confirmed in subsequent 
observations, HeH$^+$, a molecule of cosmological interest, remains undetected. To date, 
various models, including equilibrium and time-dependent models,
have been constructed to study the chemistry in envelopes and clumps, but none of them is fully 
satisfying ({\cite[Kimura 2012, and reference therein]{ki12}}).

Early observational studies of molecules in PNe mainly focused on H$_2$ and CO 
(\cite[e.g. Zuckerman \& Gatley 1988, Huggins et al. 1996]{zg88,hu96}).
Later on, millimetre observations were undertaken to trace the chemical variations of PNe
(\cite[Bachiller et al. 1997, Josselin \& Bachiller 2003]{ba97,jb03}), which indicated 
an appreciable enrichment of HCO$^+$ and CN radicals in PNe with respect to AGB envelopes.
Infrared observations have proven to be useful for searching for new small molecules
(\cite[Cernicharo et al. 1997, Liu et al. 1997]{liu97}) and fullerenes
(\cite[Cami et al. 2010, Garc{\'{\i}}a-Hern{\'a}ndez et al. 2010]{ca10,gh10}) in PNe.
The list of molecules in PNe continues to lengthen with more sensitive radio observations 
(\cite[e.g. Zhang et al. 2008, Tenenbaum et al. 2009, Edwards \& Ziurys 2013,2014, Schmidt \& Ziurys 2016]{zhang08,te2009,ez13,ez14,sz16}) and the launch of the {\it Herschel} Space Observatory
(\cite[e.g. Bujarrabal et al. 2012,  van de Steene, this volume]{bu12}).
Table\,\ref{tab1} shows the molecules discovered so far in PNe. Although most of them
are simple molecules, a big gap exists between the molecules with 5 and 60 atoms. It is
reasonable to hypothesize that complex molecules ($\ge6$ atoms) are generally present 
in PNe, and it is worthwhile to pursue future observations to search for the missing molecules.
The aim of this paper is to selectively review the studies of molecules observed in PNe, mostly focusing on the results published in the last few years.




\section{Small gas-phase molecules}
{\underline{\it H$_2$ and CO}}.
The number of PNe harboring H$_2$ has recently been considerably increased by
the UWISH2 survey (\cite[Froebrich et al. 2015, Gledhill, this  volume]{fr15}),
in which 284 extended H$_2$ sources are identified as PNe or PN candidates.
H$_2$ emission appears stronger in bipolar PNe with equatorial rings 
(\cite[Marquez-Lugo et al. 2013, and reference therein]{ma13}). 
\cite[Manchado et al. (2015 and this volume)]{ma15} presented
high-resolution H$_2$ images  showing that the H$_2$ emission in the central torus appears 
clumpy rather than uniform, and suggested that those clumps might eventually populate
the ISM and contribute to the baryonic dark matter. 
Based on the {\it Herschel} observations of an evolved PN NGC\,6720, 
\cite[van Hoof et al.(2010)]{van10} concluded that H$_2$ could be reformed on dust grains
after the central star enters the cooling track. So far more than 80 PNe have been 
detected with CO (\cite[Huggins et al. 1996,2005]{hu96,hu05}).
Early observations revealed large variations in the CO properties from object to object, with 
 young PNe exhibiting larger CO mass than evolved ones (\cite[Bujarrabal 2006]{bu06}).
However, recent observations show that the CO abundance does not significantly vary 
during PN evolution (\cite[Edwards et al. 2014]{ed14}), suggesting that substantial
amount of molecules can be expelled from PNe into the ISM.

{\underline{\it H$_2$O and OH}}. When CEs enter the PN phase, H$_2$O and OH masers will 
rapidly disappear ($<10^3$\,yrs), and thus can trace young PNe. To date, five PNe with H$_2$O 
and seven with OH maser emission have been confirmed, among which only two exhibit both
maser emission (\cite[Gomez et al. 2015a, Uscanga et al. 2012, Qiao et al. 2016]{go15a,us12,qi16}).
\cite[Gomez et al. (2015b)]{go15b} discovered the first ``water fountain'' PN
that was speculated to be the youngest PN known so far. H$_2$O and OH infrared emission has been detected in 
the O-rich PN NGC\,6302 (\cite[Bujarrabal et al. 2012]{bu12}) and the
C-rich PN NGC\,7027 (\cite[Liu et al. 1996, Wesson et al. 2010]{liu96,we10}).
A possible explanation for the presence of oxygen compounds in C-rich environments
is that the PN might have experienced a transition from O- to C-rich chemistry 
following a helium shell flash. However, a more plausible  explanation is that
they origin from an ongoing photo-induced process in PDRs
(\cite[Santander-Garc{\'{\i}}a et al. 2012]{sa12}).

\begin{figure}[t!]
\begin{center}
 \includegraphics[width=3.7in]{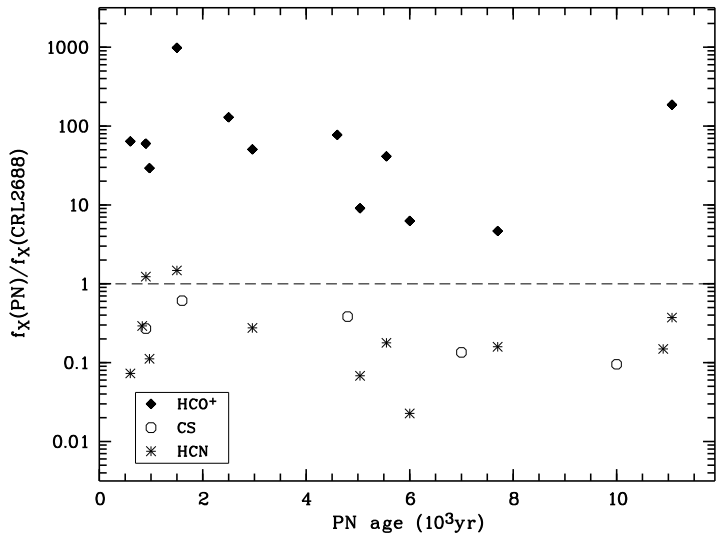} 
 \caption{Molecular abundance (wrt H$_2$) ratios between PNe and the PPN CRL\,2688 versus PN ages.
The data are taken
from \cite[Bachiller et al. (1997), Zhang et al. (2008,2013), Edwards et al. (2014), and Schmidt \& Ziurys (2016)]{ba97,zhang08,zhang13,ed14,sz16}. 
The dash line represents equivalent abundances.
}
  \label{fig1}
\end{center}
\end{figure}

{\underline{\it OH$^+$, CO$^+$, and HCO$^+$}}.
The three cationic molecules can be formed in PDRs through the path
${\rm CO} \xrightarrow{hv} {\rm O} \xrightarrow{{\rm H}^+} {\rm O}^+ \xrightarrow{{\rm H}_2} {\rm OH}^+ \xrightarrow{{\rm H}_2} {\rm H_2O}^+ \xrightarrow{{\rm H}_2} {\rm H_3O}^+ \xrightarrow{{\rm e}} {\rm OH} \xrightarrow{{\rm C}^+} {\rm CO}^+ \xrightarrow{{\rm H}_2} {\rm HCO}^+$. For the first time,
OH$^+$ was discovered in both young and evolved PNe by the {\it Herschel} Space Observatory
(\cite[Aleman et al. 2014, Etxaluze et al. 2014]{al14,et14}). 
In  evolved PNe, direct photoionization of CO  provides an alternative route to produce
 CO$^+$. \cite[Bell et al. (2007)]{bell07} detected CO$^+$ in a sample of PNe with HCO$^+$ emission. 
The correlation between the CO$^+$ and HCO$^+$ line intensities
 provides support for  the formation route of HCO$^+$ from  CO$^+$. 
HCO$^+$  is a commonly detected molecule in PNe.
\cite[Bachiller et al. (1997)]{ba97} found that HCO$^+$ abundance
increases by two orders of magnitude during the PPN-PN transition. Recent observations
suggest that the HCO$^+$ abundance is not closely
related to the ages of PNe (Fig.\,\ref{fig1}).
In highly HCO$^+$ or OH$^+$ enriched PNe, soft X-ray from the center may play an important role
in chemistry, which results in abundant H$_3^+$ and then enhances the reaction ${\rm CO} \xrightarrow{{\rm H}_3^+} {\rm HCO}^+$ or ${\rm O} \xrightarrow{{\rm H}_3^+} {\rm OH}^+$.

{\underline{\it CH and CH$^+$}}. Methylidyne radical and its cation, the important parent
species leading to the formation of CN and HCN, have been detected in 
the infrared spectrum of NGC\,7027 
(\cite[Cernicharo et al. 1997, Liu et al. 1997, Wesson et al. 2010]{cer97,liu97,we10}).  They can be formed through the
endothermic reaction ${\rm C}^+ + 0.4{\rm eV}\xrightarrow{{\rm H}_2} {\rm CH}^+$
followed by ${\rm CH}^+ \xrightarrow{{\rm H}_2} {\rm CH_2}^+ \xrightarrow{{\rm e}} {\rm CH}$. 
The  activation barrier for the formation of CH$^+$ may be overcome by
the energy carried by shocks or vibrationally excited H$_2$ in PDRs.

{\underline{\it N-bearing molecules}. 
CN has a relatively high abundance in PNe (\cite[Bachiller et al. 1997]{ba97}), which can be 
attributed to the photodissociation of HCN. Chemical models predict that, even considering
the shielding of clumps, significant destruction of HCN in PNe occurs
at a timescale of $\sim10^4$\,yrs (\cite[Redman et al. 2003]{red03}). However, this was not supported by recent observations
that suggested an approximately constant HCN abundance during the PN evolution
(\cite[Fig.\,\ref{fig1}, Schmidt \& Ziurys 2016]{sz16}). HNC, a less stable isomer of HCN, has generally
increasing abundance in PNe with respect to PPNe (\cite[Bachiller et al. 1997]{ba97}). However, the HNC/HCN
isomer ratio in NGC\,7027 appears abnormally low, which has not been well understood. 
HC$_3$N and NH$_3$, unlike those in the AGB and PPNe, are rarely detected 
in PNe. HC$_3$N was detected in NGC\,7027 with low abundance (\cite[Zhang et al. 2008]{zhang08}).
NGC\,6302 shows NH$_3$ emission (\cite[Bujarrabal et al. 2012]{bu12}). The two molecules
can either be the remnant of circumstellar chemistry, or be reformed through neutral reactions in 
the dense, warm post-shock layers of PNe.
The abundance of N$_2$H$^+$ in NGC\,7027 is extremely high (\cite[Josselin \& Bachiller 2003, Zhang et al. 2008]{jb03,zhang08}). 
This might be due to the strong X-ray radiation which enhances the reaction ${\rm N}_2 \xrightarrow{{\rm H}_3^+} {\rm N_2H}^+$.
This molecule was detected in NGC\,6537 and  NGC\,6302 as well 
(\cite[Edwards \& Ziurys 2013, Hebden 2014]{ez13,heb14}).

{\underline{\it S- and Si-bearing molecules}}. PNe are characterized with a lack of  S- and Si-bearing 
molecules, presumably because of depletion onto dust grains. CS rotational lines are 
typically strong in C-rich AGB 
envelopes and PPNe, but never detected in C-rich PNe. Only a few O-rich PNe exhibit weak CS emission
 (\cite[Fig.\,\ref{fig1}, Edwards et al. 2014, Smith et al. 2016]{ed14,sm16}),
probably suggesting that oxidation environments are favorable for 
sulfur in grains to be released into the gas phase.
HCS$^+$ and SiC$_2$ were only tentatively detected in NGC\,7027 
and NGC\,6302, respectively (\cite[Zhang et al. 2008, Hebden 2014]{zhang08,heb14}).
The other uncommon molecules SO, SO$_2$, and SiO were detected in  M\,2-28 (\cite[Edwards \& Ziurys 2014]{ez14}).
SiO, SO, and NS were revealed in the dense molecular regions of NGC\,6302 
by the SMA and ALMA observations (\cite[Hebden 2014, Santander-Garc{\'{\i}}a et al. 2016]{heb14,sg16}).

{\underline{\it C$_2$H, $c$-C$_3$H$_2$, and H$_2$CO}}. The three compounds are present in young and
evolved PNe (\cite[e.g. Woods \& Nyman 2005, Tenenbaum et al. 2009]{wn05, te09}), suggesting that they can
survive intense radiation fields.
Initiating with acetylene, C$_2$H and the cyclic molecule  $c$-C$_3$H$_2$ can be formed through photochemistry.
However, \cite[Fuente et al. (2003)]{fu03} found that the large $c$-C$_3$H$_2$ abundance in NGC\,7027  is beyond
the predictions of gas-phase chemistry models, and  might arise from
the photodestruction of large aromatic/aliphatic compounds. Although the formation of 
H$_2$CO in PDRs might result from the radical reaction between O and CH$_3$,
it is unclear whether dust surface reactions
(${\rm CO} \xrightarrow{{\rm H}} {\rm HCO} \xrightarrow{{\rm H}} {\rm H_2CO}$) followed
by photodesorption also play a role.

\section{Complex organic species}

\begin{figure}[t!]
\begin{center}
\hspace*{0.6cm}
 \includegraphics[width=2.1in]{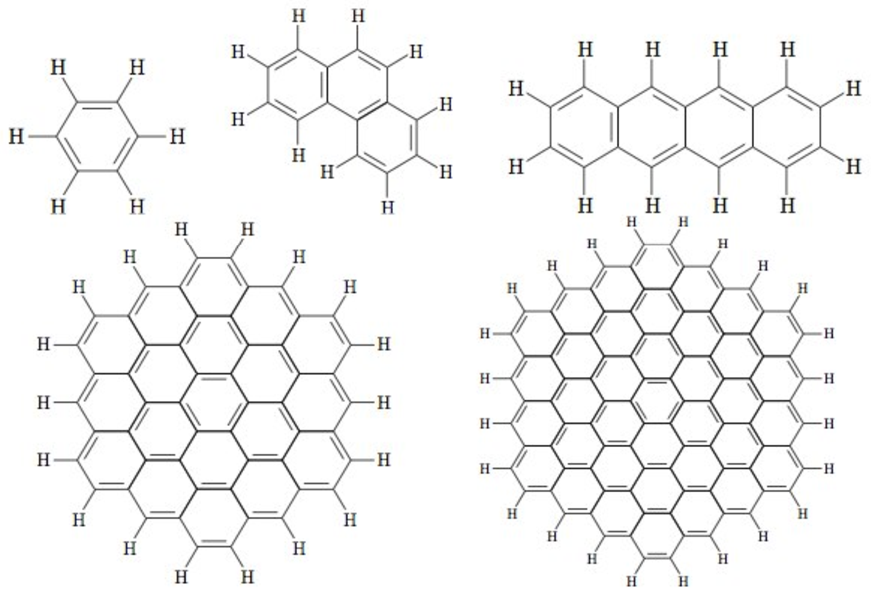}
\hspace*{0.1cm}
 \includegraphics[width=2.8in]{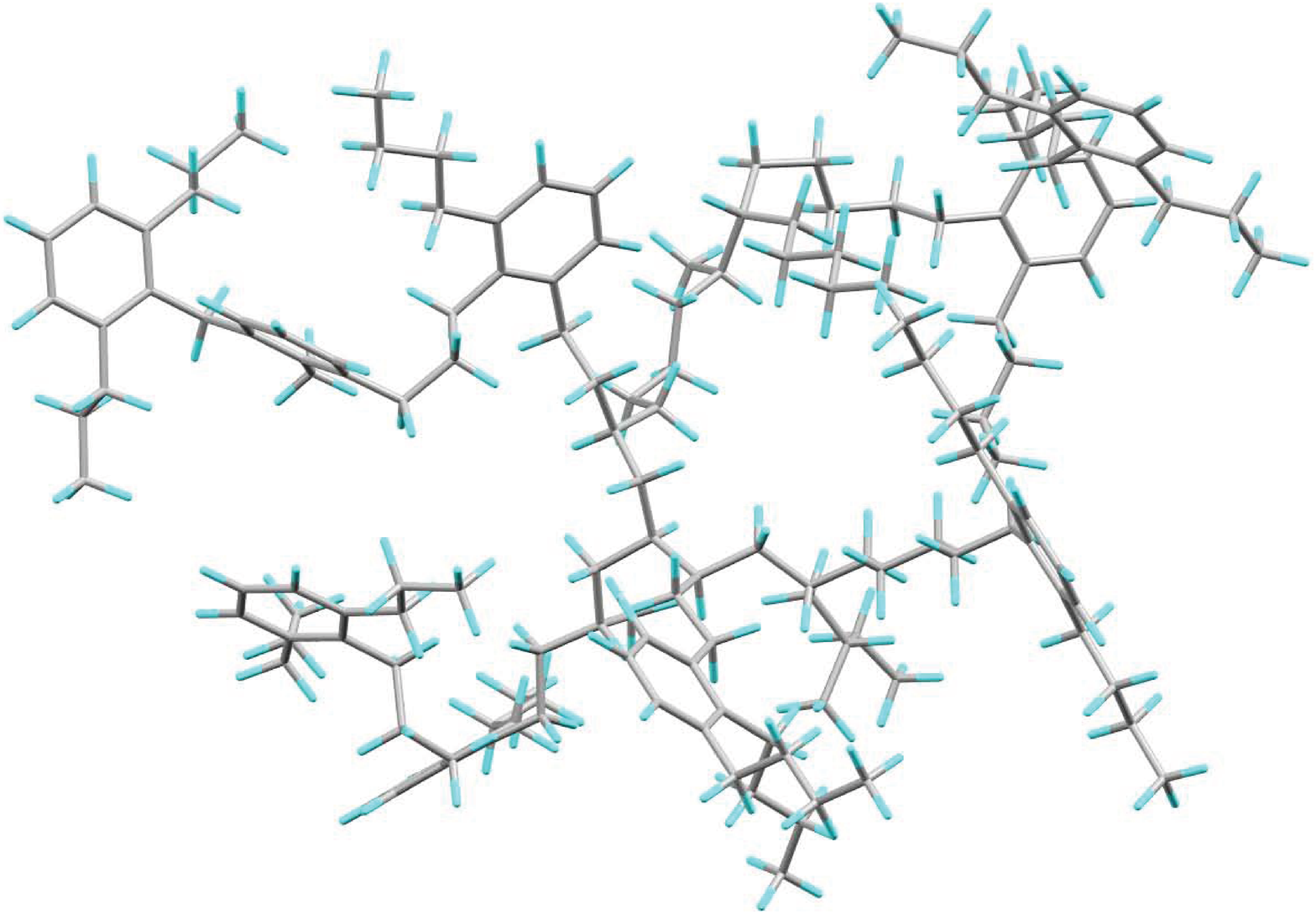}
 \caption{Schematic chemical structures of PAHs (left) and MAON (right).
}  \label{fig2}
\end{center}
\end{figure}

The presence of complex  organics can be inferred from the unidentified infrared 
emission (UIE) bands that mainly manifest themselves as  aromatic features at 
3.3, 6.2, 7.7, 8.6, and 11.3\,$\mu$m, aliphatic features at 3.4, and 6.9\,$\mu$m, and
plateau emission around 8, 12, and 17\,$\mu$m (\cite[Tielens 2008, Kwok 2011, and reference therein]{ti08,kwok11}).
UIE bands have been observed in a variety of astronomical environments, and their fluxes show that
the carrier might be the most abundant molecules in the ISM. Both low- and high-excitation PNe
exhibit UIE bands, indicating that their carrier is a significant nebular component.
A wide variety of materials has been proposed as potential carriers of the UIE bands, among
which polycyclic aromatic hydrocarbon (PAH) is the most commonly accepted one
(\cite[see Peeters 2013, for a recent review]{pe13}).  However, the PAH hypothesis
is not without problems  (\cite[Kwok \& Zhang 2013, Zhang \& Kwok 2015]{kz13,zk15}).
\cite[Kwok \& Zhang (2011)]{kz11} invoked an alternative model, mixed
aromatic/aliphatic organic nanoparticles (MAONs), to interpret the UIE phenomenon.
Fig.\,\ref{fig2} shows the chemical structures of the PAHs and MAON.
The spectral analysis shows that the aromatic/aliphatic component ratio increases
during the transition from PPN to PN, suggesting that the aliphatic component may
be processed to more stable aromatic rings by ultraviolet photons. The destruction of MAONs
can provide a top-down route for the formation of smaller molecules. Nevertheless, no
consensus has been reached concerning the UIE carrier. The MAON model was recently questioned 
(\cite[e.g. Li \& Draine 2012, Yang et al. 2016]{ld12,yang16}).

Recent observations  of NGC\,7027 provide support for the formation of the UIE carrier through 
grain-grain collisions in post-shock regions (\cite[Lau et al. 2016]{lau16}).
UIE bands were also discovered in O-rich PNe. Infrared images of O-rich 
PNe show that UIE primarily originates from PDRs associated with the outer edge of central tori
or dense knots, suggesting that the UIE carrier can be formed through a
bottom-up process following the  photodissociatoin of CO
(\cite[Guzman-Ramirez et al. 2014, Cox et al. 2016]{gr14,cox16}).

\section{Fullerene-related compounds}

\begin{figure}[t!]
\begin{center}
 \includegraphics[width=2.0in]{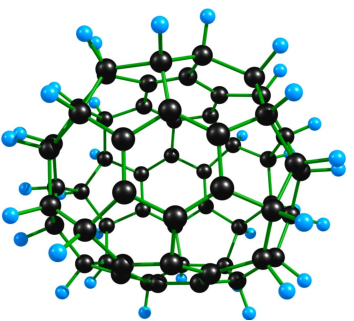}
\hspace*{0.5cm}
 \includegraphics[width=2.5in]{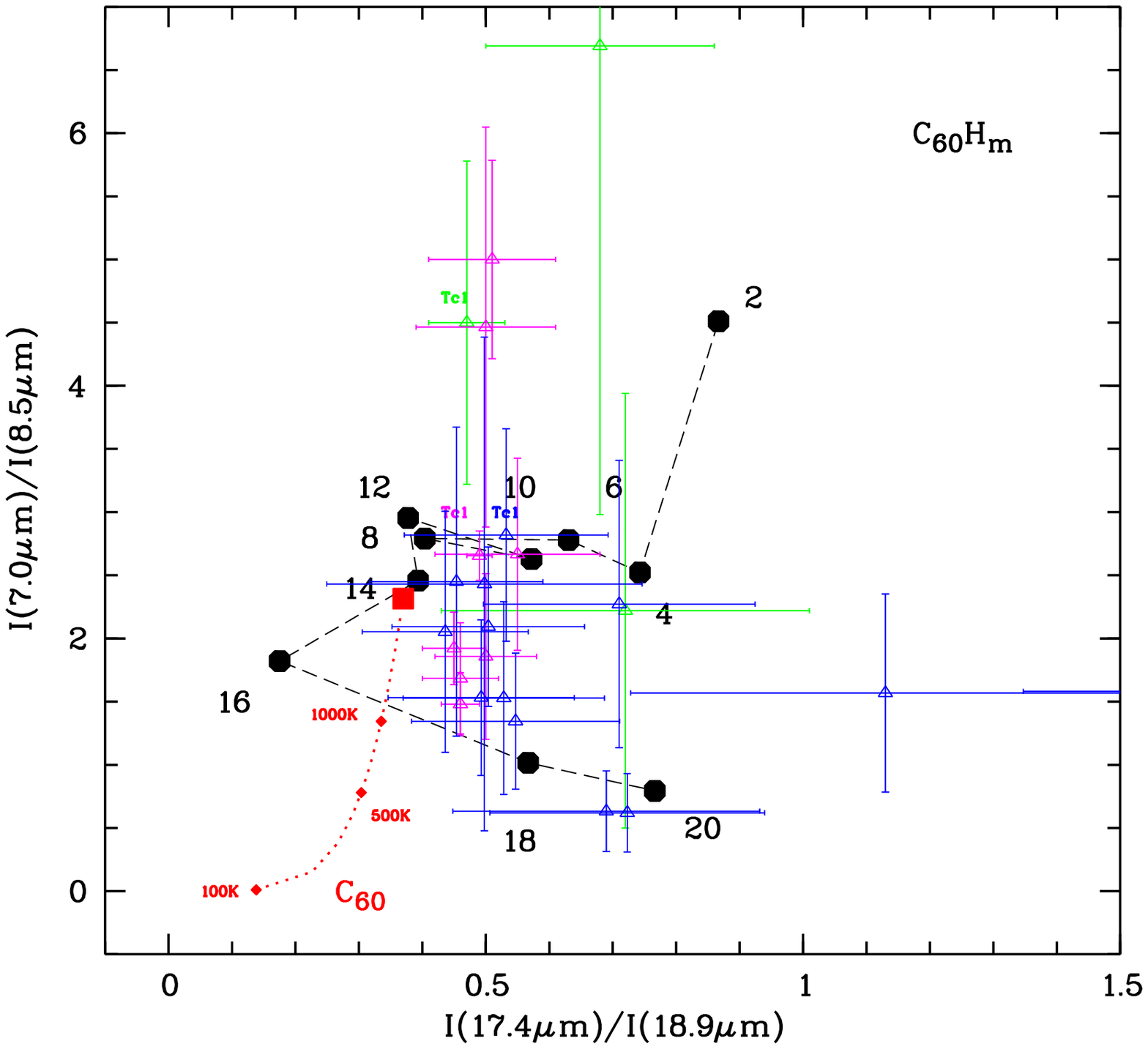}
 \caption{
{\it Left}: Optimized structures of a C$_{60}$H$_{\rm 36}$ isomer.
{\it Right}: Theoretical intrinsic strength ratios of
the four C$_{60}$H$_{\rm m}$ backbone vibrations (filled circles connected
by dashed line, with the m values marked). 
The triangles with error bars are the observed
values taken from \cite[Bernard-Salas et al. (2012), Garc{\'{\i}}a-Hern{\'a}ndez et al. (2012),
and Otsuka et al.(2014)]{bc12,gar12,ots14}.
The dotted curve represents the predicted ratios based upon  the thermal excitation 
model of C$_{60}$.
}  \label{fig3}
\end{center}
\end{figure}

Given its high stability, C$_{60}$, the best-known fullerene member, has
been long believed to exist throughout circumstellar and interstellar environments
(\cite[Kroto et al.1985]{kroto}). Because of high symmetry, 
the vibrational modes of C$_{60}$ are degenerated into four individual
infrared features, 
making the astronomical search possible.
Soon after the first detection in a young PN (\cite[Cami et al. 2010]{ca10}),
C$_{60}$  was discovered in various astronomical environments,
suggesting that it could survive  in very harsh conditions, though the formation route
is under  a controversial debate (\cite[see Cami et al. 2011 and this volume, Zhang \& Kwok 2013, Zhang et al. 2016, for the details]{ca11,zk13,zk16}).  So far, 11 Galactic PNe
and 11 Magellanic Cloud PNe have been detected with C$_{60}$ (\cite[Garc{\'{\i}}a-Hern{\'a}ndez et al. 2012, Sloan et al. 2014, Otsuka et al. 2016]{gar12,sl14,ots16}).
The detection rate is higher in more metal-poor environments.  The presence 
of C$_{60}$ in a PPN suggests that it can be formed within a timescale of $<10^3$\,yrs 
(\cite[Zhang \& Kwok 2011]{zk11}). Given its rather low ionization potential,
C$_{60}$ presumably  exists largely in the cation form in the ISM. The
notion of C$_{60}^+$  as one of the DIB carriers
(\cite[Foing \& Ehrenfreund 1994]{f94}) was recently verified  by gas-phase experimental 
data, that is, five near-infrared DIBs were convincingly assigned to C$_{60}^+$
(\cite[Campbell et al. 2015,2016, Walker et al. 2015]{c15,wb15,c16}).
A few DIBs were found to be enhanced in C$_{60}$-containing nebulae
(\cite[Iglesias-Groth \& Esposito 2013, D{\'{\i}}az-Luis et al. 2015]{ig13,dg15}),
suggesting that they might have a circumstellar origin and probably be related to fullerenes
or derivatives.  Given their chemical activity and physical stability, fullerene-related compounds are good candidates for the DIB carrier 
(\cite[see Omont 2016, for a comprehensive discussion]{om16}).
Future studies should examine the spectra of
fullerene compounds including hydrogenated fullerenes (C$_{60}$H$_{\rm m}$), C$_{60}$ adducts, endofullerenes, hererofullerenes, buckyonions, and so on.

If mixed with H atoms, C$_{60}$ can be readily hydrogenated into C$_{60}$H$_{\rm 36}$
(Fig.\,\ref{fig3}) in laboratory environments. Thus C$_{60}$H$_{\rm m}$ is highly
likely to be present in PNe, and may contribute to the formation of H$_2$
by dehydrogenation (\cite[Cataldo \& Iglesias-Groth 2009]{ci09}).  A tentative
detection of C$_{60}$H$_{\rm m}$ in a C$_{60}$-containing PPN has been presented
by \cite[Zhang \& Kwok (2013)]{zk13}. However, two C$_{60}$-rich PNe 
do not exhibit any detectable C-H feature (\cite[D{\'{\i}}az-Luis et al. 2016]{dg16}).
Theoretical investigation indicated that with slight hydrogenation, 
the four C$_{60}$-backbone vibration features at 7.0, 8.5, 17.4, and 18.9\,$\mu$m
are still visible but diminish their intensities in 
different degrees (\cite[Zhang et al. 2016]{zk16}). Therefore, C$_{60}$-poor PNe might be 
more ideal targets for the search of C$_{60}$H$_{\rm m}$.
The excitation mechanism of C$_{60}$  bands is still unclear, hindering
their use as a probe of PN environments (\cite[Brieva et al. 2016]{bg16}).
As shown in Fig.\,\ref{fig3}, the observed intensity ratios are too dispersed to accord 
with the excitation model characterized with one single free parameter 
(i.e. fluorescence or thermal excitation models 
governed by the photon field energy and temperature, respectively), 
but can be well interpreted in terms of hydrogenation (Zhang et al., in preparation).  This might provide a
circumstantial evidence for the presence of  C$_{60}$H$_{\rm m}$ in PNe.

\section{Conclusion}

A wealth of observations have demonstrated that molecules are an essential
component in PNe. Even large molecules can be formed and/or survive during
the PN evolution. The complex molecules blown from PNe into the ISM
might contribute to the DIBs and seed the chemistry of molecular clouds.
However, the complete picture of the chemistry in PNe is still far from 
clear. Systematic molecular line surveys of PNe are relatively scarce.
Many molecules are discovered in only one or a few
PNe. There is an absence of thorough understanding of the connection between
the molecular composition and the properties of PNe. It is expected that 
new and important insights will be offered by the next generation instruments. 
Searching for fullerene-related compounds in PNe might provide
crucial clues toward solving the DIB enigma.  Further theoretical and experimental studies are 
badly needed.

\acknowledgements
I thank the SOC for inviting me to do this review talk. 
I am indebted to my colleagues who have closely collaborated with me
in the study of circumstellar molecules, including Sun Kwok, Jun-ichi 
Nakashima, and Seyedabdolreza Sadjadi.
Financial support for this work was provided by the Research Grants Council of the Hong Kong under grants HKU7073/11P and HKU7062/13P.

\end{document}